\documentclass[aps,prl,twocolumn,groupedaddress,showpacs]{revtex4}

\usepackage{bm}
\usepackage{graphicx}

\begin{document}

\title{Evolution of a pairing-induced pseudogap from the superconducting gap of (Bi,Pb)$_2$Sr$_2$CuO$_6$}

\author{
	K. Nakayama,$^1$
	T. Sato,$^1$
	Y. Sekiba,$^1$
	K. Terashima,$^1$
	P. Richard,$^2$
	T. Takahashi,$^{1,2}$
	K. Kudo,$^3$
	N. Okumura,$^3$
	T. Sasaki,$^3$
	and N. Kobayashi$^3$}

\affiliation{$^1$Department of Physics, Tohoku University, Sendai 980-8578, Japan}
\affiliation{$^2$WPI Research Center, Advanced Institute for Materials Research, Tohoku University, Sendai 980-8577, Japan}
\affiliation{$^3$Institute for Materials Research, Tohoku University, Sendai 980-8577, Japan}

\date{\today}

\begin{abstract}
We have performed an ultrahigh-resolution angle-resolved photoemission spectroscopy study of slightly-overdoped (Bi,Pb)$_2$Sr$_2$CuO$_6$ to elucidate the origin of pseudogap.  By using a newly developed xenon-plasma light source, we determined the comprehensive momentum and temperature dependences of the superconducting gap and the pseudogap.  We found that the antinodal pseudogap persists far above the superconducting transition temperature and is smoothly connected to the nodal gap.  The characteristic temperature of the pseudogap scales well with the superconducting-gap size irrespective of the momentum location.  The present experimental results point to the pairing origin of the pseudogap.
\end{abstract}


\maketitle

The anomalies of various physical properties in underdoped high-$T_{\rm c}$ (transition temperature) cuprate superconductors above $T_{\rm c}$ are related to the opening of an energy gap at the Fermi level ($E_{\rm F}$) known as a pseudogap (PG).  The PG has been observed in both spin and charge excitations, and regarded as a key ingredient to understand the mechanism of high-$T_{\rm c}$ superconductivity.  There are two different proposals to explain the PG; one is associated with the formation of local incoherent pairs above $T_{\rm c}$ (precursor pairing) \cite{Emery} and the other is not directly related to the superconductivity and rather competes with superconductivity \cite{Kampf, Li, Chakravarty}.  The former is based on the experimental results from angle-resolved photoemission spectroscopy (ARPES) \cite{Ding, Loeser, NormanNature} and tunneling spectroscopy \cite{Renner, Kugler}, which demonstrate a smooth evolution of the superconducting (SC) gap into the PG across $T_{\rm c}$, and the overall $d$-wave symmetry of PG \cite{Kanigel, Valla}.  The latter scenario is based on ARPES \cite{Tanaka, Kondo, Ma, Kondo2}, tunneling spectroscopy \cite{Boyer}, and Raman scattering \cite{Tacon} studies, which reported two different types of energy gaps; a SC gap which opens below $T_{\rm c}$ in the nodal region and an antinodal PG whose magnitude shows a $T$-independence across $T_{\rm c}$.  These apparently different proposals cause a critical disagreement on the nature of PG.  One of the main reasons to cause such a controversy is the lack of experimental data on the systematic $T$- and $k$-dependences of the PG and the SC gap in a wider parameter range.  Since it has been reported that the two-gap-like behavior is significantly enhanced in single-layered low-$T_{\rm c}$ cuprates \cite{Kondo, Ma, Kondo2, Terashima}, Bi$_2$Sr$_2$CuO$_6$ (Bi2201) provides a good opportunity to investigate these issues.

In this Letter, we report an ARPES study of the PG and the SC gap in slightly-overdoped Bi2201.  By using a newly developed xenon (Xe)-plasma light source \cite{SoumaRSI} which enables ultrahigh energy- and momentum-resolved measurements, we have revealed a smooth connection between the SC gap and the PG in the $k$ space as well as across $T_{\rm c}$, in contrast to the observation of two-energy scales in underdoped Bi2201 \cite{Kondo, Ma, Kondo2}.  We also found that the PG has a $d$-wave ground state.  These experimental results strongly suggest that the observed PG is caused by the precursor pairing.  We also provide a possible clue to settle the controversy on the origin of the PG.

\begin{figure}[!t]
\begin{center}
\includegraphics[width=3in]{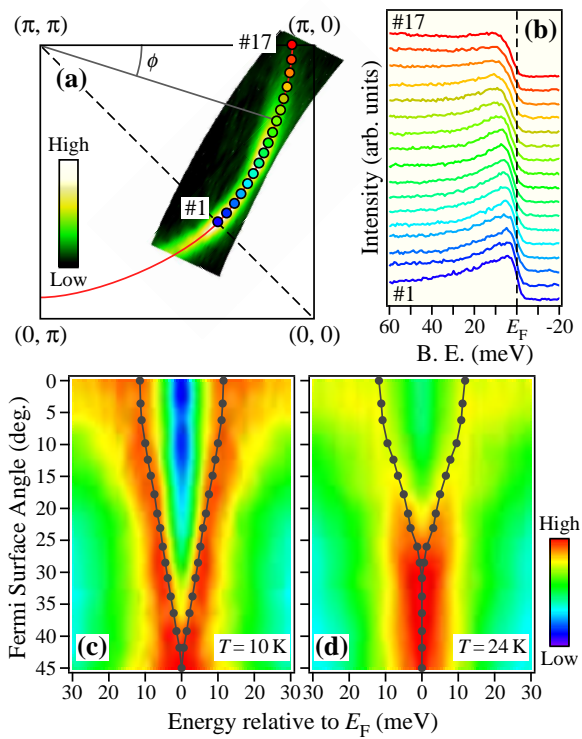}
\end{center}
\caption{
(color online) (a) Plot of ARPES intensity at $E_{\rm F}$ as a function of two-dimensional wave vector measured at 10 K with Xe I line ($h$$\nu$ = 8.437 eV) for Pb-Bi2201 ($T_{\rm c}$ = 21 K).  The intensity was obtained by integrating the spectral intensity within $\pm$15 meV with respect to $E_{\rm F}$, and normalized to the photon flux.  Colored circles show the position of the Fermi vectors ($k_{\rm F}$) determined by the minimum-gap locus method \cite{NormanNature}.  It is noted that the $k_{\rm F}$'s are located on the normal-state FS (red line).  (b) ARPES spectra measured at 10 K at various $k_{\rm F}$ points.  Coloring of the spectra is the same as that of circles in (a).  Spectra $\sharp$1 and $\sharp$17 were measured along the nodal (0, 0)-($\pi$, $\pi$) and the antinodal ($\pi$, 0)-($\pi$, $\pi$) cuts, respectively.  The spectral intensity is normalized to that of the peak maximum.  (c) and (d) Symmetrized ARPES-spectral intensity at $k_{\rm F}$ points as a function of FS angle ($\phi$) measured at $T$ = 10 K and 24 K, respectively.  Black circles show the SC-gap size $\Delta$ determined by fitting the symmetrized spectra with the phenomenological gap function \cite{NormanPRB}.
}
\end{figure}

High-quality single crystals of slightly-overdoped (Bi,Pb)$_2$Sr$_2$CuO$_6$ (Pb-Bi2201) were grown by the floating-zone method \cite{Kudo1,Kudo2}.  The hole concentration was controlled by annealing the sample under nitrogen atmosphere at high temperature.  The onset $T_{\rm c}$ of samples determined by the magnetic susceptibility measurement is 21 K.  Ultrahigh-resolution ARPES measurements were performed using a VG-SCIENTA SES2002 photoemission spectrometer with a newly developed Xe-plasma light source \cite{SoumaRSI}.  We used one of the Xe I lines ($h$$\nu$ = 8.437 eV) to excite photoelectrons.  Since electrons near $E_{\rm F}$ excited by the Xe I line ($h$$\nu$ = 8.437 eV) have a relatively long escape depth (20-40 \AA) compared with that (5-10 \AA) of the He I ($h$$\nu$ = 21.218 eV), the Xe I ARPES spectrum reflects mostly the bulk electronic structure.  The energy and angular (momentum) resolutions were set at 2-3 meV and 0.2$^{\circ}$ (0.004 \AA$^{-1}$), respectively.  We cleaved samples under an ultrahigh vacuum of 2$\times$10$^{-11}$ Torr to obtain a clean and fresh sample surface for measurements.  The Fermi level ($E_{\rm F}$) of samples was referenced to that of a gold film evaporated onto the sample substrate.

Figure 1(a) shows the ARPES intensity at $E_{\rm F}$ plotted as a function of the two-dimensional wave vector for Pb-Bi2201.  We find a large holelike Fermi surface (FS) centered at the ($\pi$, $\pi$) point of Brillouin zone, consistent with previous ARPES studies with higher-energy photons \cite{Sato, Kondo}.  We clearly observe a SC gap at 10 K at various Fermi-vector ($k_{\rm F}$) points [Fig. 1(b)], although the size of SC gap is much smaller than that of Bi2212 \cite{Mesot, Lee}.  Along the nodal cut (spectrum $\sharp$1), the leading-edge midpoint (LEM) is located at $E_{\rm F}$ while it is gradually shifted toward higher binding energy on approaching the antinodal cut ($\sharp$17), showing the anisotropic nature of the SC gap.  As shown in Fig. 1(c), where the intensity of symmetrized ARPES spectra at 10 K is plotted as a function of FS angle ($\phi$), the SC gap clearly exhibits a $d_{x^2-y^2}$-like behavior with a point node at $\phi$ = 45$^{\circ}$.  The observed gap size $\Delta$ is well fitted by the $d$-wave gap function including the second higher harmonic term \cite{Mesot} as $\Delta$($\phi$) = $\Delta$$_{max}$(0.89cos(2$\phi$)+0.11cos(6$\phi$)) with $\Delta$$_{max}$ = 11.3 meV.  As shown in Fig. 1(d), we find that a similar energy gap opens even above $T_{\rm c}$.  The gap magnitude around the antinode (0$^{\circ}$ $<$ $\phi$ $<$ 10$^{\circ}$) is almost the same as that below $T_{\rm c}$, while the gap around the node (30$^{\circ}$ $<$ $\phi$ $<$ 45$^{\circ}$) is totally absent above $T_{\rm c}$, indicating the coexistence of an antinodal PG and a nodal Fermi arc \cite{NormanNature} above $T_{\rm c}$.

\begin{figure*}[!t]
\begin{center}
\includegraphics[width=6.6in]{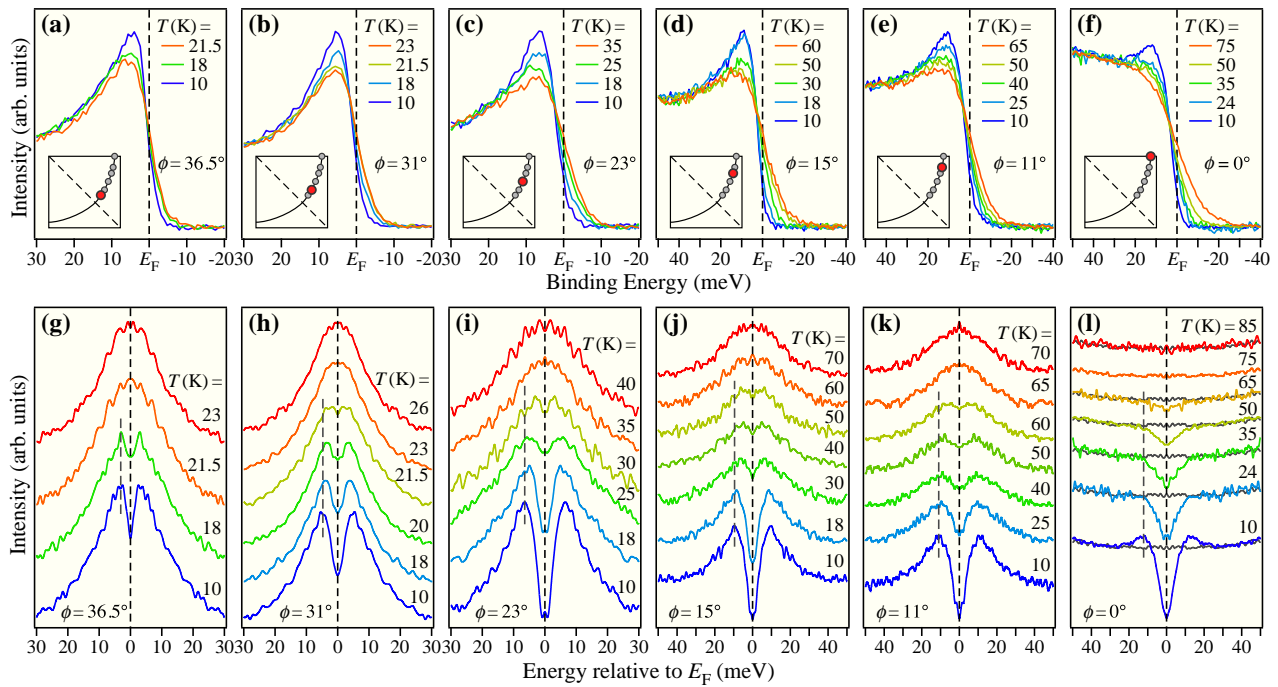}
\end{center}
\caption{
(color online) (a)-(f) Temperature dependence of ARPES spectra measured at various $k_{\rm F}$ points shown by a red circle in each inset.  (g)-(l), same as (a)-(f) but symmetrized with respect to $E_{\rm F}$.  The gray dashed line denotes the peak position at 10 K.  In (l), the 75 K-spectrum (gray line) is superimposed on each spectrum for comparison in (l).
}
\end{figure*}

Figure 2 displays the $T$-dependence of ARPES spectra measured at various $k_{\rm F}$ points.  As shown in panels (a)-(f), the LEM, which is located around $E_{\rm F}$ above $T_{\rm c}$, is gradually shifted toward higher binding energy with decreasing $T$, showing the SC-gap opening.  We also find that a sharp peak emerges at 12 meV at the antinode below $T_{\rm c}$ [panel (f)] while it disappears above $T_{\rm c}$, indicating that the peak is a SC coherence peak.  In Figs. 2(g)-(l), we plot the corresponding $T$-dependence of the symmetrized ARPES spectrum at each $k_{\rm F}$ point.  Around the nodal region [Figs. 2(g) and (h)], the energy gap looks to close almost at $T_{\rm c}$ (21 K) as evident from the presence of a single peak at $E_{\rm F}$ above $T_{\rm c}$.  On the other hand, in the $k$ region away from the node [panels (i)-(l)], we find two peaks even above $T_{\rm c}$ due to the PG opening.  At the antinode [Fig. 2(l)], the energy scale of the PG is identical to that of the SC gap, and the gap is gradually filled in on increasing $T$.  The PG finally vanishes at $T^*$ $\sim$ 75 K ($>$ 3$T_{\rm c}$) in good agreement with the previous tunneling experiment of overdoped Bi2201 \cite{Kugler}.  This result indicates that the electronic states of Bi2201 suffer strong influences from the PG even in the overdoped region in contrast to Bi2212 which shows the PG opening only in a narrow temperature range above $T_{\rm c}$ \cite{Ding, NormanNature, NormanPRB}.  Interestingly, in the $k$ region between the node and the antinode [off-nodal region, panels (i)-(k)], the peak approaches $E_{\rm F}$ on increasing $T$, demonstrating that unlike the antinodal gap, the off-nodal energy gap $closes$, but $is$ $not$ $filled$ $in$.

\begin{figure}[!t]
\begin{center}
\includegraphics[width=3in]{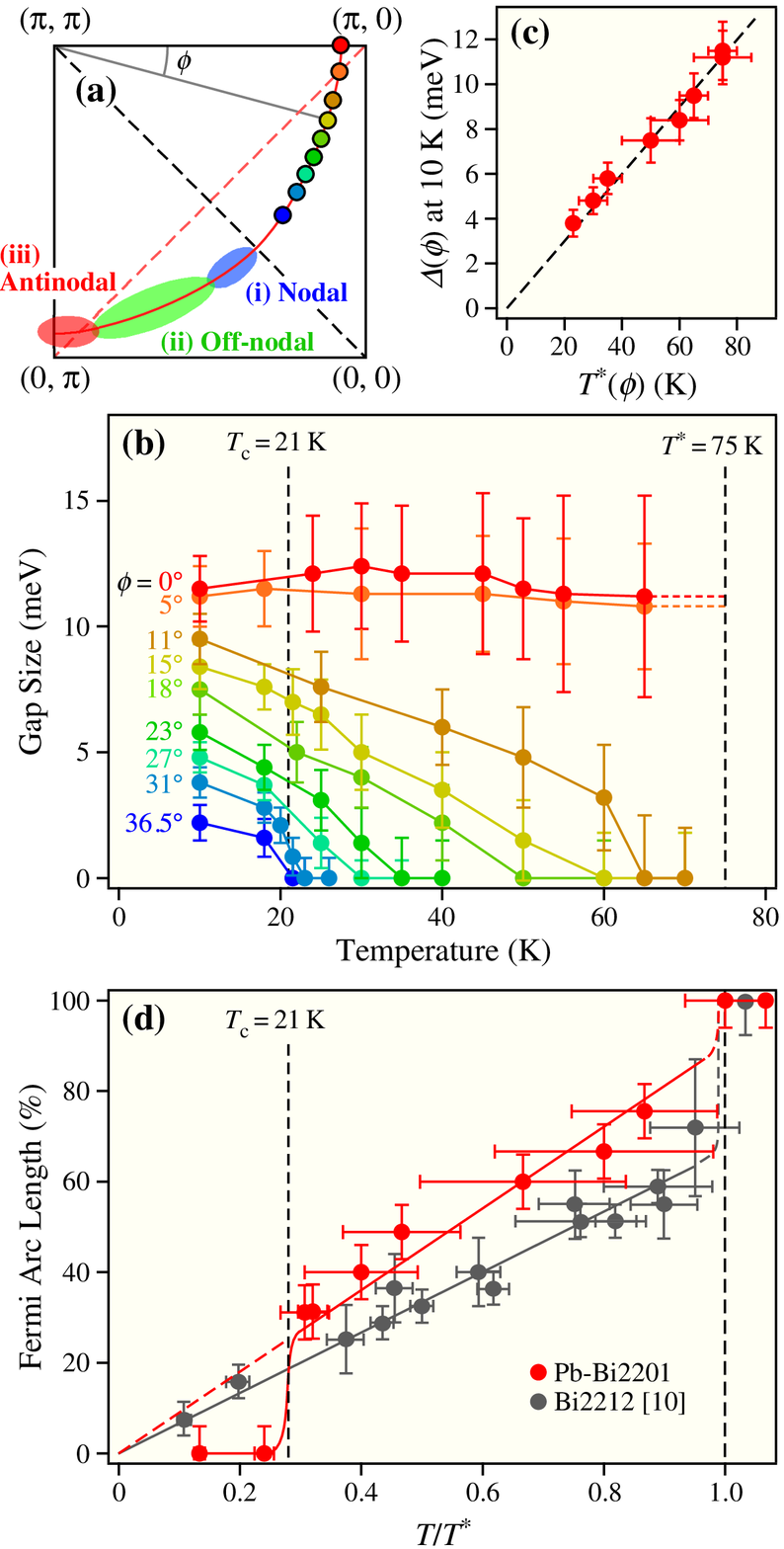}
\end{center}
\caption{
(color online) (a) Schematic diagram of three characteristic $k$ regions (i)-(iii) where the energy gap shows a characteristic $T$-dependence.  In the nodal region (i), the gap opens only below $T_{\rm c}$.  In the off-nodal region (ii), a PG exists above $T_{\rm c}$ and it gradually $closes$ on increasing $T$, while in the antinodal region (iii), the PG $is$ $filled$ $in$ by keeping the gap-size constant.  (b) $T$-dependence of the energy gap for different $\phi$ determined by fitting the symmetrized spectra with the phenomenological BCS function \cite{NormanPRB} convoluted with the instrumental resolution.  (c) Gap size $\Delta$ at 10 K plotted as a function of $T^*$ for various $\phi$ values.  The dashed line is a guide to highlight the linear relation.  (d) Length of the Fermi arc as a function of $T$/$T^*$ (red circles), estimated by the data in (b).  The dashed and solid lines indicate that the arc length becomes zero when $T$/$T^*$ approaches zero.  The $T$-dependence of the Fermi arc length in Bi2212 \cite{Kanigel} is also shown for comparison (gray circles).
}
\end{figure}

To estimate quantitatively the energy-gap size, we fitted the symmetrized spectra by the phenomenological BCS spectral function convoluted with the instrumental energy resolution, as applied in previous studies \cite{NormanPRB, Kanigel, Lee}.  In Fig. 3(b), we plot the $T$-dependence of the energy gap at various $k_{\rm F}$ points indicated by circles in Fig. 3(a).  While the nodal gap ($\phi$ = 31$^{\circ}$ and 36.5$^{\circ}$) closes around $T_{\rm c}$, the off-nodal PG ($\phi$ = 11$^{\circ}$-27$^{\circ}$) persists even above $T_{\rm c}$ and closes at $T^*$ higher than $T_{\rm c}$.  Around the antinode ($\phi$ = 0$^{\circ}$ and 5$^{\circ}$), the gap value is always about 12 meV and does not show a marked $T$-dependence.  It is inferred that there are three characteristic $k$ regions categorized by the behavior of PG [see Fig. 3(a)]: (i) the PG is absent, (ii) the PG closes above $T_{\rm c}$, and (iii) the PG is filled in above $T_{\rm c}$.  It is noted that while the characteristic difference in the $T$-dependence of the nodal and antinodal gaps has been reported previously \cite{Kanigel, NormanPRB, Lee}, it was not clear whether these gaps are smoothly connected in the $k$ space.  It is remarked that the $T$-dependence of the energy gap has no discontinuity between the nodal and the off-nodal region, suggesting the same origin for the SC gap and the off-nodal PG.  Moreover, although the $T$-dependence of the PG in the off-nodal region is qualitatively different from that in the antinodal region, the two PG may have a related origin because the $T^*$ value at the off-node is smoothly connected to that at the antinode.  This argument is supported by the experimental data in Fig. 3(c), which show a linear relation for $\Delta$($\phi$) at 10 K $vs$ $T^*$($\phi$), indicating that $T^*$($\phi$) scales with the SC-gap size rather than the PG size just above $T_{\rm c}$.  These results, which have not been revealed in the previous ARPES experiments, cannot be reconciled by the simple summation of two different gaps (nodal and antinodal gaps).

In Fig. 3(d), we plot the length of the Fermi arc as a function of $T$/$T^*$.  Below $T_{\rm c}$, the Fermi-arc length is zero since the SC gap has a point node.  On increasing $T$, it suddenly jumps at $T_{\rm c}$ and monotonically increases up to $T^*$.   When we extrapolate the Fermi-arc length at $T_{\rm c}$ $<$ $T$ $<$ $T^*$ to $T$ = 0 K, the arc length approaches zero (see red dashed line) in good agreement with that in Bi2212 (gray circles) \cite{Kanigel}.  This demonstrates that the PG and the SC gap are both $d$-wave-like in overdoped Bi2201, and implies that the $d$-wave ground state of the PG is a universal feature of hole-doped cuprates \cite{Kanigel, Valla}.

Now we discuss the origin of the observed PG.  As seen in Figs. 1(d) and 3(b), the PG extends to the nodal region ($\phi$ $\sim$ 30$^{\circ}$) above $T_{\rm c}$, leaving a short Fermi arc around the nodal region.  It is noted that the Fermi arc (30$^{\circ}$ $<$ $\phi$ $<$ 45$^{\circ}$) is located far away from the hot spot ($\phi$ $\sim$ 5$^{\circ}$) defined as the intersection of the normal-state FS and the antiferromagnetic zone boundary.  The spin-density-wave (SDW) \cite{Kampf} and the $d$-density-wave scenarios \cite{Chakravarty} have predicted that a PG opens around the hot spot and subsequently causes a considerable difference between the minimum gap locus and the normal-state FS.  However, since such a difference is not observed as seen in Fig. 1(a), both scenarios are unlikely to account for the PG.  The absence of well-nested parallel segments in the FS near the antinode [Fig. 1(a)] is also unfavorable to a CDW origin of PG \cite{Li} in contrast to underdoped Ca$_{2-x}$Na$_x$CuO$_2$Cl$_2$ \cite{Shen}.  One of plausible explanation for PG is the precursor pairing above $T_{\rm c}$ \cite{Emery}, because (i) the PG has the same $d$-wave symmetry as the SC gap, (ii) the nodal, off-nodal and antinodal gaps are smoothly connected with each other, (iii) no discontinuity of the energy-gap size is observed across $T_{\rm c}$, and (iv) $T^*$ is simply characterized by the SC-gap size irrespective of $k$.  Although our data demonstrate that the PG and the SC gap in slightly overdoped Bi2201 have the same energy scales and the same $k$ anisotropy indicative of a single $d$-wave gap, it has been reported that there are two energy scales at the antinode in more underdoped samples \cite{Kondo, Ma, Kondo2, Terashima, Shi}.  One is a low-energy peak assigned as the SC coherence peak and the other is a broad feature observed at higher energy.  The latter has been often regarded as a hallmark of the presence of a PG which competes with the superconductivity \cite{Kondo, Ma, Kondo2}.  Since the present comprehensive ARPES study clearly shows that such a broad feature is absent even in the doping range where the superconductivity is observed, it is inferred that the competing PG is not a unique feature of the SC samples.  The ARPES spectra of underdoped samples are more naturally explained in terms of the superimposition of a low energy scale PG associated with the precursor pairing and a larger energy scale PG due to a density-wave formation \cite{Kampf, Chakravarty, Li, Shen, Ma, Matsui}.  Thus, the complicated behavior of the PG in Bi2201 is well understood in terms of the presence of two different types of PGs.

In summary, we reported ultrahigh-resolution ARPES results on slightly-overdoped Pb-Bi2201 ($T_{\rm c}$ = 21 K).  We have determined the precise $k$- and $T$-dependences of the SC gap and PG by using a low-energy Xe-plasma light source, and found that the PG smoothly evolves from the SC gap.  The observed novel scaling behavior of $T^*$ with the SC-gap size as well as the $T$ dependence of the Fermi-arc length strongly suggests that the origin of the observed PG is the pairing rather than the competing order.

We thank H. Ding and S. Souma for useful discussions.  This work was supported by grants from JST-CREST and MEXT of Japan.  K. N. and K. T. thank JSPS for financial support.


\begin{thebibliography}{99}

\bibitem{Emery} V. J. Emery $et$ $al$., Nature {\bf 374}, 434 (1995).
\bibitem{Kampf} A. Kampf and J. R. Schrieffer, Phys. Rev. B {\bf 41}, 6399 (1990).
\bibitem{Chakravarty} S. Chakravarty $et$ $al$., Phys. Rev. B {\bf 63}, 094503 (2001).
\bibitem{Li} J.-X. Li, C.-Q. Wu, and D.-H. Lee, Phys. Rev. B {\bf 74}, 184515 (2006).
\bibitem{Ding} H. Ding $et$ $al$., Nature {\bf 382}, 51 (1996).
\bibitem{Loeser} A. G. Loeser $et$ $al$., Science {\bf 273}, 325 (1996).
\bibitem{NormanNature} M. R. Norman $et$ $al$., Nature {\bf 392}, 157 (1998).
\bibitem{Renner} C. Renner $et$ $al$., Phys. Rev. Lett. {\bf 80}, 149 (1998).
\bibitem{Kugler} M. Kugler $et$ $al$., Phys. Rev. Lett. {\bf 86}, 4911 (2001).
\bibitem{Kanigel} A. Kanigel $et$ $al$., Nature Phys. {\bf 2}, 447 (2006).
\bibitem{Valla} T. Valla $et$ $al$., Science {\bf 314}, 1914 (2006).
\bibitem{Tanaka} K. Tanaka $et$ $al$., Science {\bf 314}, 1910 (2006).
\bibitem{Kondo} T. Kondo $et$ $al$., Phys. Rev. Lett. {\bf 98}, 267004 (2007).
\bibitem{Ma} J.-H. Ma $et$ $al$., Phys. Rev. Lett. {\bf 101}, 207002 (2008).
\bibitem{Kondo2} T. Kondo $et$ $al$., Nature {\bf 457}, 296 (2009).
\bibitem{Boyer} M. C. Boyer $et$ $al$., Nature Phys. {\bf 3}, 802 (2007).
\bibitem{Tacon} M. L. Tacon $et$ $al$., Nature Phys. {\bf 2}, 537 (2006).
\bibitem{Terashima} K. Terashima $et$ $al$., Phys. Rev. Lett. {\bf 99}, 017003 (2007).
\bibitem{SoumaRSI} S. Souma $et$ $al$., Rev. Sci. Instrum. {\bf 78}, 123104 (2007).
\bibitem{Kudo1} K. Kudo $et$ $al$., J. Phys. Soc. Jpn. {\bf 75}, 124710 (2006).
\bibitem{Kudo2} K. Kudo $et$ $al$., Physica C {\bf 426-431}, 251 (2005).
\bibitem{Sato} T. Sato $et$ $al$., Phys. Rev. B {\bf 64}, 054502 (2001).
\bibitem{Mesot} J. Mesot $et$ $al$., Phys. Rev. Lett. {\bf 83}, 840 (1999).
\bibitem{Lee} W. S. Lee $et$ $al$., Nature {\bf 450}, 81 (2007).
\bibitem{NormanPRB} M. R. Norman $et$ $al$., Phys. Rev. B {\bf 57}, R11093 (1998).
\bibitem{Shen} K. M. Shen $et$ $al$., Science {\bf 307}, 901 (2005).
\bibitem{Shi} M. Shi $et$ $al$., Phys. Rev. Lett. {\bf 101}, 047002 (2008).
\bibitem{Matsui} H. Matsui $et$ $al$., Phys. Rev. B {\bf 75}, 224514 (2007).

\end{thebibliography}
\end{document}